\documentclass[11pt]{article}
\usepackage{epsfig}
\usepackage{amsmath,amsfonts}

\begin{document}

\title{\textbf{Remarks on the dynamical mass generation in confining Yang-Mills theories}}
\author{S.P. Sorella\thanks{%
sorella@uerj.br}{\ }\footnote{Work supported by FAPERJ, Funda{\c
c}{\~a}o de Amparo {\`a} Pesquisa do Estado do Rio de Janeiro,
under the program {\it Cientista do Nosso Estado}, E-26/151.947/2004.} \\
{\small {\textit{Departamento de F\'{\i}sica Te\'orica,}}}\\
{\small {\textit{Instituto de F\'{\i}sica, Universidade do Estado
do Rio de
Janeiro,}}}\\
{\small {\textit{\ Rua S\~{a}o Francisco Xavier 524, 20550-013
Maracan\~{a}, }}}{\small {\textit{Rio de Janeiro, Brazil.}}}}
\date{}
\maketitle

\begin{abstract}
The dynamical mass generation for gluons is discussed in Euclidean
Yang-Mills theories supplemented with a renormalizable mass term.
The mass parameter is not free, being determined in a
self-consistent way through a gap equation which obeys the
renormalization group. The example of the Landau gauge is worked
out explicitly at one loop order. A few remarks on the issue of
the unitarity are provided.
\end{abstract}

\newpage

\section{Introduction}

In the last years many efforts have been done to put forward the
idea that gluons might acquire a mass through a dynamical
mechanism. These efforts have led to a considerable amount of
evidence, obtained through theoretical and phenomenological
studies \cite
{Cornwall:1981zr,Greensite:1985vq,Stingl:1985hx,Lavelle:1988eg,Gubarev:2000nz,Gubarev:2000eu,
Verschelde:2001ia, Kondo:2001nq,Kondo:2001tm,Dudal:2003vv,
Browne:2003uv,Ellwanger:2002sj,Dudal:2003gu,
Dudal:2003by,Aguilar:2002tc, Dudal:2004rx,
Browne:2004mk,Gracey:2004bk,Parisi:1980jy,Field:2001iu,Szczepaniak:2003mr,Li:2004te},
as well as from lattice simulations \cite
{Amemiya:1998jz,Boucaud:2001st,Boucaud:2002nc,Boucaud:2005rm,
Langfeld:2001cz,
Alexandrou:2000ja,Alexandrou:2001fh,Bornyakov:2003ee,RuizArriola:2004en,Suzuki:2004dw,Chernodub:2005gz}.
Many aspects related to the dynamical gluon mass generation
deserve a better understanding. This is the case, for example, of
the unitarity of the resulting theory, a highly nontrivial topic,
due to the confining character of $QCD$. \newline
\newline
Needless to say, the unitarity of the $S$ matrix is a fundamental property
of the spectrum of a quantum field theory. It expresses the conservation of
the probability of the amplitudes corresponding to the various scattering
processes among the excitations of the spectrum. \newline
\newline
In a nonconfining theory, the first step in the construction of the $S$
matrix is the introduction of the so-called $\left| in\right\rangle $ and $%
\left| out\right\rangle $ Fock spaces characterizing the asymptotic behavior
of the physical states before, $t\rightarrow -\infty $, and after, $%
t\rightarrow +\infty $, a scattering process. The $S$-matrix is thus defined
as the unitary operator which interpolates between the spaces $\left|
in\right\rangle $ and $\left| out\right\rangle $, namely
\begin{equation}
\left| in\right\rangle =S\;\left| out\right\rangle \;.  \label{s1}
\end{equation}
The relation of this equation with the Green's functions of the
theory is provided by the $LSZ$ formalism. A key ingredient of
this formalism is the introduction of the asymptotic fields,
$\varphi _{in}$ and $\varphi _{out}$, describing the asymptotic
behavior of the interacting fields $\varphi $, according to
\begin{eqnarray}
\left. \varphi \right| _{t\rightarrow -\infty } &=&Z^{1/2}\varphi _{in}\;,
\label{s2} \\
\left. \varphi \right| _{t\rightarrow +\infty } &=&Z^{1/2}\varphi _{out}\;.
\nonumber
\end{eqnarray}
The asymptotic fields $\varphi _{in}$ and $\varphi _{out}$ allow us to
define the creation and annihilation operators $\left( a_{in}^{\dagger
},\;a_{in}\right) $ and $\left( a_{out}^{\dagger },\;a_{out}\right) $, from
which the Fock spaces $\left| in\right\rangle $ and $\left| out\right\rangle
$ are obtained. The entire construction relies on the possibility that the
asymptotic fields can be consistently introduced. \newline
\newline
In a confining theory like $QCD$, the quanta associated with the
basic fields of the theory, \textit{i.e.} the gluon field $A_{\mu
}^{a}$ and the quark fields $\psi $, $\overline{\psi }$, cannot be
observed as free particles, due to color confinement. The physical
spectrum of the theory is made up by colorless bound states of
quarks and gluons giving rise, for instance, to barions, mesons
and glueballs. This implies that the asymptotic Fock spaces
$\left| in\right\rangle $,  $\left| out\right\rangle $ of the
theory have to be defined through suitable operators from which
the physical spectrum of the excitations is constructed. Of course, the $S$%
-matrix describing the scattering amplitudes among the excitations
of the physical spectrum of the theory has to be unitary. Although
intuitively simple and easily understandable, this framework is
far beyond our present capabilities. An operational definition of
the gauge invariant colorless operators defining the physical
spectrum of the excitations and a well defined set of rules to
evaluate their scattering amplitudes are not yet at our disposal.
\newline
\newline
Quantized Yang-Mills theories are described by the Faddeev-Popov Lagrangian%
\footnote{%
From now on we shall consider pure Yang-Mills theory in the Euclidean
space-time.}
\begin{equation}
S=S_{YM}+S_{gf}=\int d^{4}x\;\left( \frac{1}{4}F_{\mu \nu }^{a}F_{\mu \nu
}^{a}\;+b^{a}\partial _{\mu }A_{\mu }^{a}+\overline{c}^{a}\partial _{\mu
}\left( D_{\mu }c\right) ^{a}\right) \;,  \label{s3}
\end{equation}
here taken in the Landau gauge. The field $b^{a\text{ }}$in expression $%
\left( \ref{s3}\right) $ is the Lagrange multiplier enforcing the Landau
gauge condition, $\partial _{\mu }A_{\mu }^{a}=0$, while $\overline{c}^{a}$,
$c^{a}$ stand for the Faddeev-Popov ghosts. The action $\left( \ref{s3}%
\right) $ is renormalizable to all orders of perturbation theory and
displays color confinement\footnote{%
Although we still lack a definite proof of color confinement, it is
doubtless that pure Yang-Mills theory, as given in eq.$\left( \ref{s3}%
\right) $, displays such a phenomenon.}. Furthermore, thanks to
the asymptotic freedom, the gauge field $A_{\mu }^{a}$ behaves
almost freely at very high energies, where perturbation theory is
reliable. However, at low energies, the coupling constant
increases and the effects of color confinement cannot be
neglected. We do have thus a good understanding of the properties
of the field $A_{\mu }^{a}$ at high energies, whereas it becomes
more and more difficult to have a clear picture of $A_{\mu }^{a}$,
and of the whole theory, as the energy decreases. We might thus
adopt the point of
view of starting with a renormalizable action built up with a gauge field $%
A_{\mu }^{a}$ which accommodates the largest possible number of
degrees of freedom. This would amount to start with a quantized
massive Yang-Mills action
\begin{equation}
S_{m}=S_{YM}+S_{gf}+S_{mass}\;,  \label{n1}
\end{equation}
where $S_{mass}$ is a suitable mass term for the gauge field
$A_{\mu }^{a}$. The best choice for $S_{mass}$ would be a gauge
invariant, renormalizable local mass term. However, no local
renormalizable gauge invariant mass term built up with gauge
fields only is at our disposal. Nevertheless, it might be worth
reminding that, recently, a consistent framework for the nonlocal
gauge invariant mass operator
\begin{equation}
\mathcal{O}(A)=-\frac{1}{2}\int d^{4}xF_{\mu \nu }^{a}\left[
\left( D^{2}\right) ^{-1}\right] ^{ab}F_{\mu \nu }^{b}\;.
\label{gm}
\end{equation}
has been achieved \cite{Capri:2005dy}. More precisely, the
nonlocal operator $\left( \ref{gm} \right)$ can be cast in local
form by means of the introduction of a suitable set of additional
fields. The resulting local theory displays the important property
of being multiplicatively renormalizable \cite{Capri:2005dy}.
\\\\Though, for the time being, we give up of the requirement of
the gauge invariance. This will enable us to present our analysis
with the help of a relatively simple example. Therefore, as
possible mass term we shall take
\begin{equation}
S_{mass}=\frac{1}{2}m^{2}\int d^{4}xA_{\mu }^{a}A_{\mu }^{a}\;,  \label{n2}
\end{equation}
so that
\begin{equation}
S_{m}=\int d^{4}x\;\left( \frac{1}{4}F_{\mu \nu }^{a}F_{\mu \nu }^{a}\;+%
\frac{1}{2}m^{2}A_{\mu }^{a}A_{\mu }^{a}+b^{a}\partial _{\mu }A_{\mu }^{a}+%
\overline{c}^{a}\partial _{\mu }\left( D_{\mu }c\right) ^{a}\right) \;.
\label{s4}
\end{equation}
Expression $\left( \ref{s4}\right) $ provides an example of a
massive nonabelian gauge theory which is renormalizable to all
orders of perturbation theory \cite {Dudal:2002pq}, while obeying
the renormalization group equations (RGE).
\newline
\newline
A few remarks are now in order:

\begin{itemize}
\item  The amplitudes corresponding to the scattering processes among gluons
and quarks display now a violation of the unitarity. This can be understood
by noting that the inclusion of the mass term $m^{2}A_{\mu }^{a}A_{\mu }^{a}$
gives rise to a $BRST$ operator which is not nilpotent. However, as shown in
\cite{Dudal:2002pq}, it is still possible to write down suitable
Slavnov-Taylor identities which ensure that the massive theory $\left( \ref
{s4}\right) $ is renormalizable to all orders of perturbation theory.
Moreover, if sufficiently small, this violation of the unitarity might not
be in conflict with the confining character of the theory. Otherwise said,
since gluons are not directly observable, we could allow for a gauge field $%
A_{\mu }^{a}$ with the largest possible number of degrees of
freedom, provided the renormalizability is preserved and one is
able to recover the results of the massless case at very high
energies.

\item  This framework would be useless if the value of the mass
parameter $m$ would be free, meaning that we are introducing a new
arbitrary parameter in the theory, thereby changing its physical
meaning. A different situation is attained by demanding that the
mass parameter is determined in a self-consistent way as a
function of the coupling constant $g$. This can be obtained by
requiring that the mass $m$ in eq.$\left( \ref{s4}\right) $ is a
solution of a suitable gap equation. In other words, even if the
mass $m$ is included in the starting gauge-fixed theory, it does
not play the role of a free parameter, as it is determined once
the quantum effects are properly taken into account. Here, we rely
on the lack of an exact description of a confining Yang-Mills
theory at low energies. We start then with the largest possible
number of degrees of freedom compatible with the renormalizability
requirement and fix the mass parameter through the gap equation.
If the resulting value of $m$ will be small enough, one can argue
that the unitarity is violated by terms which become less and less
important as the energy of the process increases, so that the
amplitudes of the massless case are in practice recovered at very
high energies. The present set up might thus provide a different
characterization of the aforementioned phenomenon of the dynamical
gluon mass generation, which has already been successfully
described in \cite
{Cornwall:1981zr,Verschelde:2001ia,Dudal:2003vv,Browne:2003uv,Dudal:2003gu,
Dudal:2003by,Dudal:2004rx,Browne:2004mk}. In the next section, the
gap equation for the mass $m$ will be discussed.
\end{itemize}

\section{The gap equation for the mass parameter $m$}

The gap equation for the mass parameter $m$ is obtained by requiring that
the vacuum functional $\mathcal{E}$ defined by
\begin{equation}
e^{-V\mathcal{E}}=\int \left[ D\Phi \right] \;e^{-\left( S_{m}+V\eta (g)%
\frac{m^{4}}{2}\right) }\;,  \label{s5}
\end{equation}
where $V$ is the Euclidean space-time volume, obeys a minimization condition
with respect to the mass $m$, \textit{i.e.} the value of the mass $m$ is
determined by demanding that it corresponds to the minimum of the vacuum
functional $\mathcal{E}$, namely
\begin{equation}
\frac{\partial \mathcal{E}}{\partial m^{2}}=0\;.  \label{s6}
\end{equation}
Equation $\left( \ref{s6}\right) $ is the gap equation for the mass
parameter $m$. The quantity $\eta (g)$ in eq.$\left( \ref{s5}\right) $ is a
dimensionless parameter whose loop expansion
\begin{equation}
\eta (g)=\eta _{0}(g)+\hbar \eta _{1}(g)+\hbar ^{2}\eta _{2}(g)+....
\label{sd}
\end{equation}
accounts for the quantum effects related to the renormalization of the
vacuum diagrams in the massive case. The parameter $\eta (g)$ can be
obtained order by order by requiring that the vacuum functional $\mathcal{E}$
obeys the renormalization group equations (RGE)
\begin{equation}
\mu \frac{d\mathcal{E}}{d\mu }=0\;,  \label{s7}
\end{equation}
meaning that $\mathcal{E}$ is independent from the renormalization scale $%
\mu $, as it will be explicitly verified in the next section. Equation $%
\left( \ref{s7}\right) $ expresses an important property of the vacuum
functional $\mathcal{E}$. We also remark that a term of the kind of $\eta
m^{4}$ in eq.$\left( \ref{s5}\right) $ has been already obtained\footnote{%
See eq.(6.22) of Sect.VI of \cite{Cornwall:1981zr}.} in \cite
{Cornwall:1981zr} in the evaluation of the vacuum energy of Yang-Mills
theories when gluons are massive. \newline
\newline
The gap equation equation $\left( \ref{s7}\right)$ can be given a
simple interpretation. Due to the lack of an exact description of
Yang-Mills theories at low energies, we have adopted the point of
view of starting with a renormalizable massive action, as given in
eq.$\left( \ref{s4}\right) $. As far as the mass parameter $m$ is
free, expression $\left( \ref{s4}\right) $ can be interpreted as
describing a family of massive models, parametrized by $m$. For
each value of $m$ we have a specific renormalizable model.
Moreover, as the introduction of a mass term has an energetic
coast, we might figure out that, somehow, the dynamics will select
precisely that model corresponding to the lowest energetic coast,
as expressed by the gap equation $\left( \ref{s7}\right) $.
\\\\Before starting with explicit calculations let us summarize
our point of view:

\begin{itemize}
\item  Since gluons are not directly observable, we allow for a gauge field $%
A_{\mu }^{a}$ with the largest number of degrees of freedom compatible with
the requirement of renormalizability.

\item  This amounts to start with a renormalizable massive action, as given
in eq.$\left( \ref{s4}\right) $. However, the mass parameter $m$ is
determined in a self-consistent way by imposing the minimizing condition $%
\left( \ref{s6}\right) $ on the vacuum functional $\mathcal{E}$.

\item  Also, it is worth observing that, in the case of the
massive model of eq.$\left( \ref{s4}\right) $, a non vanishing
solution, $m_{sol}^{2}\neq 0 $, of the gap equation $\left(
\ref{s6}\right) $ implies the existence of a non vanishing
dimension two gluon condensate $\left\langle A_{\mu }^{a}A_{\mu
}^{a}\right\rangle $. In fact, differentiating equation $\left(
\ref{s5}\right) $ with respect to $m^{2}$ and setting
$m^{2}=m_{sol}^{2}$, one obtains
\begin{equation}
\frac{1}{2}\left\langle A_{\mu }^{a}A_{\mu }^{a}\right\rangle =-\eta
m_{sol}^{2}{\ .}  \label{s8}
\end{equation}
\end{itemize}

\section{Evaluation of the vacuum functional $\mathcal{E}\;$at one loop order
}

In the case of pure $SU(N)$ Yang-Mills theories, for the vacuum functional $%
\mathcal{E}$ we have
\begin{equation}
e^{-V\mathcal{E}}=\int \left[ D\Phi \right] \;e^{-\left( S_{m}+V\eta \frac{%
m^{4}}{2}\right) }\;,  \label{ym1}
\end{equation}
with $S_{m}$ given by expression $\left( \ref{s4}\right) $, namely
\begin{equation}
S_{m}=\int d^{4}x\;\left( \frac{1}{4}F_{\mu \nu }^{a}F_{\mu \nu }^{a}\;+%
\frac{1}{2}m^{2}A_{\mu }^{a}A_{\mu }^{a}+b^{a}\partial _{\mu }A_{\mu }^{a}+%
\overline{c}^{a}\partial _{\mu }\left( D_{\mu }c\right) ^{a}\right) \;.
\label{ym2}
\end{equation}
As it has been proven in \cite{Dudal:2002pq}, the massive action $\left( \ref
{ym2}\right) $ is multiplicatively renormalizable to all orders of
perturbation theory. In particular, for the mass renormalization we have
\cite{Dudal:2002pq}
\begin{eqnarray}
g_{0} &=&Z_{g}g\;,  \nonumber \\
A_{0} &=&Z_{A}^{1/2}A  \nonumber \\
m_{0}^{2} &=&Z_{m^{2}}m^{2}\;,  \nonumber \\
Z_{m^{2}} &=&Z_{g}Z_{A}^{-1/2}\;,  \label{ym3}
\end{eqnarray}
from which the running of the mass $m^{2}$ is easily deduced
\begin{equation}
\mu \frac{\partial m^{2}}{\partial \mu }=-\gamma _{m^{2}}m^{2}\;,
\label{ymm3}
\end{equation}
with
\begin{equation}
\gamma _{m^{2}}(g^{2})=\gamma _{0}g^{2}+\overline{\gamma }%
_{1}g^{4}+O(g^{6})\;,  \label{ymmm4}
\end{equation}
\begin{equation}
\gamma _{0}=\frac{35}{6}\frac{N}{16\pi ^{2}}\;,\;\;\;\;\;\;\;\overline{%
\gamma }_{1}=\frac{449}{24}\left( \frac{N}{16\pi ^{2}}\right) ^{2}\;.
\label{ymmm5}
\end{equation}
Also
\begin{equation}
\beta (g^{2})=\overline{\mu }\frac{\partial g^{2}}{\partial \overline{\mu }}%
=-2\left( \beta _{0}g^{4}+\beta _{1}g^{6}+O(g^{8})\right) \;,  \label{yy5}
\end{equation}
\begin{equation}
\beta _{0}=\frac{11}{3}\frac{N}{16\pi ^{2}}\;,\;\;\;\;\;\;\beta _{1}=\frac{34%
}{3}\left( \frac{N}{16\pi ^{2}}\right) ^{2}\;.  \label{yy51}
\end{equation}
In order to obtain the parameter $\eta $ at one-loop order, it is useful to
note that expression $\left( \ref{ym1}\right) $ can be written in localized
form as
\begin{equation}
e^{-V\mathcal{E}}=\int \left[ D\Phi \right] \;e^{-\left( S_{m}+V\eta \frac{%
m^{4}}{2}\right) }\;=\int DJ(x)\;\delta (J(x)-m^{2})\;e^{-W(J)}\;,
\label{nym1}
\end{equation}
with
\begin{eqnarray}
e^{-W(J)} &=&\int \left[ D\Phi \right] \;e^{-S(J)\;},  \label{nym2} \\
S(J) &=&S_{YM}+S_{gf}+\int d^{4}x\left( \frac{1}{2}J(x)A_{\mu }^{a}A_{\mu
}^{a}+\frac{\eta }{2}J^{2}(x)\right) \;.  \nonumber
\end{eqnarray}
From equation $\left( \ref{nym1}\right) $ it follows that the
renormalization of the vacuum functional $\mathcal{E}$ can be achieved by
renormalizing the functional $W(J)$ in the presence of the local source $%
J(x) $, and then set $J=m^{2}$ at the end. The renormalization of the
functional $W(J)$ has been worked out at two-loops in \cite
{Verschelde:2001ia}. By simple inspection, it turns out that the parameter $%
\eta $ is related to the LCO parameter $\zeta $ of \cite{Verschelde:2001ia}
by $\eta =-\zeta $, yielding
\begin{equation}
\eta =-\frac{9}{13g^{2}}\frac{N^{2}-1}{N}-\hbar \frac{161}{52}\frac{N^{2}-1}{%
16\pi ^{2}}+O(g^{2})\;.  \label{ym5}
\end{equation}
Thus, for the vacuum functional $\mathcal{E}$ at one-loop order in the $%
\overline{MS}$ scheme, we get
\begin{equation}
\mathcal{E=}\frac{m^{4}}{2}\left( -\frac{9}{13g^{2}}\frac{N^{2}-1}{N}-\hbar
\frac{161}{52}\frac{N^{2}-1}{16\pi ^{2}}\right) +3\hbar \frac{N^{2}-1}{64\pi
^{2}}m^{4}\left( -\frac{5}{6}+\log \frac{m^{2}}{\overline{\mu }^{2}}\right)
\;,  \label{ym6}
\end{equation}
where we have introduced the factor $\hbar $ to make clear the order of the
various terms. It is useful to check explicitly that the above expression
obeys the RGE\ equations. Indeed, from eqs.$\left( \ref{ymmm4}\right) $, $%
\left( \ref{yy5}\right) $ we obtain
\begin{eqnarray}
\overline{\mu }\frac{d\mathcal{E}}{d\overline{\mu }} &=&-\hbar \gamma
_{0}g^{2}m^{4}\left( -\frac{9}{13g^{2}}\frac{N^{2}-1}{N}\right) +\hbar \frac{%
m^{4}}{2}\frac{9}{13g^{4}}\frac{N^{2}-1}{N}(-2\beta _{0}g^{4})-\hbar 6\frac{%
N^{2}-1}{64\pi ^{2}}m^{4}+O(\hbar ^{2})  \nonumber \\
&=&\hbar m^{4}\frac{N^{2}-1}{16\pi ^{2}}\left( \frac{35}{6}\right) \frac{9}{%
13}-\hbar m^{4}\frac{N^{2}-1}{16\pi ^{2}}\frac{33}{13}-\hbar 6\frac{N^{2}-1}{%
64\pi ^{2}}m^{4}+O(\hbar ^{2})  \nonumber \\
&=&\hbar m^{4}\frac{N^{2}-1}{16\pi ^{2}}\left( \frac{35}{6}\frac{9}{13}-%
\frac{33}{13}-\frac{6}{4}\right) +O(\hbar ^{2})=\hbar m^{4}\frac{N^{2}-1}{%
16\pi ^{2}}\left( \frac{105}{26}-\frac{33}{13}-\frac{3}{2}\right) +O(\hbar
^{2})  \nonumber \\
&=&\hbar m^{4}\frac{N^{2}-1}{16\pi ^{2}}\left( \frac{105-66-39}{26}\right)
+O(\hbar ^{2})=O(\hbar ^{2})\;.  \label{ym7}
\end{eqnarray}
It remains now to look for a sensible solution of the gap equation $\left(
\ref{s6}\right) $. This will be the task of the next section.

\subsection{Searching for a sensible minimum}

In order to search for a sensible solution of the gap equation $\left( \ref
{s6}\right) $, $\frac{\partial \mathcal{E}}{\partial m^{2}}=0$, we first
remove the freedom existing in the renormalization of the mass parameter by
replacing it with a renormalization scheme and scale independent quantity.
This can be achieved along the lines outlined in \cite{Dudal:2003vv} in the
analysis of the gluon condensate $\left\langle A_{\mu }^{a}A_{\mu
}^{a}\right\rangle $ within the 2PPI expansion technique. Let us first
change notation
\begin{eqnarray}
g^{2} &\rightarrow &\overline{g}^{2}\;,  \label{a1} \\
m^{2} &\rightarrow &\overline{m}^{2}\;,  \nonumber
\end{eqnarray}
and rewrite the one-loop vacuum functional as
\begin{equation}
\mathcal{E=}\frac{9}{13}\frac{N^{2}-1}{N}\frac{1}{\overline{g}^{2}}\left[ -%
\frac{\overline{m}^{4}}{2}+\frac{13}{3}\frac{N\overline{g}^{2}}{64\pi ^{2}}%
\overline{m}^{4}\left( \log \frac{\overline{m}^{2}}{\overline{\mu }^{2}}-%
\frac{113}{39}\right) \right] \;.  \label{a2}
\end{equation}
As done in \cite{Dudal:2003vv}, we introduce the scheme and scale
independent quantity $\widetilde{m}^{2}$ through the relation
\begin{equation}
\widetilde{m}^{2}=\overline{f}(\overline{g}^{2})\overline{m}^{2}\;.
\label{a3}
\end{equation}
From
\begin{equation}
\overline{\mu }\frac{\partial \overline{m}^{2}}{\partial \overline{\mu }}=-%
\overline{\gamma }_{m^{2}}(\overline{g}^{2})\overline{m}^{2}\;,  \label{a4}
\end{equation}
with
\begin{eqnarray}
\overline{\gamma }_{m^{2}}(\overline{g}^{2}) &=&\gamma _{0}\overline{g}^{2}+%
\overline{\gamma }_{1}\overline{g}^{4}+O(\overline{g}^{6})\;,  \label{a5} \\
&&  \nonumber
\end{eqnarray}
\begin{equation}
\gamma _{0}=\frac{35}{6}\frac{N}{16\pi ^{2}}\;,\;\;\;\;\;\;\;\overline{%
\gamma }_{1}=\frac{449}{24}\left( \frac{N}{16\pi ^{2}}\right) ^{2}\;,
\label{a55}
\end{equation}
we obtain the condition
\begin{equation}
\overline{\mu }\frac{\partial \overline{f}(\overline{g}^{2})}{\partial
\overline{\mu }}=\overline{\gamma }_{m^{2}}(\overline{g}^{2})\overline{f}(%
\overline{g}^{2})\;,  \label{a6}
\end{equation}
from which it follows that
\begin{equation}
\overline{\mu }\frac{\partial \widetilde{m}^{2}}{\partial \overline{\mu }}%
=0\;.  \label{a7}
\end{equation}
Equation $\left( \ref{a6}\right) $ is easily solved, yielding
\begin{eqnarray}
\overline{f}(\overline{g}^{2}) &=&(\overline{g}^{2})^{-\frac{\gamma _{0}}{%
2\beta _{0}}}\left( 1+f_{0}\overline{g}^{2}+O(\overline{g}^{4})\right) \;,
\nonumber \\
f_{0} &=&\frac{1}{2\beta _{0}}\left( \frac{\gamma _{0}}{\beta _{0}}\beta
_{1}-\overline{\gamma }_{1}\right) \;,  \label{a8}
\end{eqnarray}
where the coefficients $\beta _{0}$, $\beta _{1}\;$are given in eqs.$\left(
\ref{yy5}\right) $, $\left( \ref{yy51}\right) $. Moreover, one has to take
into account that a change of scheme entails a change in the coupling
constant $\overline{g}^{2}$, according to
\begin{equation}
\overline{g}^{2}=g^{2}(1+b_{0}g^{2}+O(g^{4}))\;.  \label{a10}
\end{equation}
The coefficient $b_{0}$ in eq.$\left( \ref{a10}\right) $ expresses the
freedom related to the choice of the renormalization scheme. It will be
fixed by demanding that the coupling constant is renormalized in such a
scheme so that the vacuum functional $\mathcal{E}$ takes the form
\begin{equation}
\mathcal{E}\left( \widetilde{m}^{2}\right) \mathcal{=}\frac{9}{13}\frac{%
N^{2}-1}{N}\frac{1}{\left( g^{2}\right) ^{1-\frac{\gamma _{0}}{\beta _{0}}}}%
\left( -\frac{\widetilde{m}^{4}}{2}+\widetilde{m}^{4}\frac{Ng^{2}}{16\pi ^{2}%
}E_{1}L\right) \;,  \label{a11}
\end{equation}
where $L$ stands for
\begin{equation}
L=\log \frac{\widetilde{m}^{2}\left( g^{2}\right) ^{\frac{\gamma _{0}}{%
2\beta _{0}}}}{\overline{\mu }^{2}}\;,  \label{a12}
\end{equation}
and $E_{1}$ is a numerical coefficient. After a simple calculation, we get
\begin{eqnarray}
\mathcal{E} &=&\frac{9}{13}\frac{N^{2}-1}{N}\frac{1}{\left( g^{2}\right) ^{1-%
\frac{\gamma _{0}}{\beta _{0}}}}\left[ -\frac{\widetilde{m}^{4}}{2}+%
\widetilde{m}^{4}\frac{13}{3}\frac{Ng^{2}}{64\pi ^{2}}\left( L-\frac{113}{39}%
+\frac{3}{13}\frac{64\pi ^{2}}{N}\left( f_{0}+\frac{b_{0}}{2}(1-\frac{\gamma
_{0}}{\beta _{0}})\right) \right) \;\right] \;.  \nonumber \\
&&  \label{a13}
\end{eqnarray}
Therefore, for $b_{0}$ one has
\begin{equation}
-\frac{113}{39}+\frac{3}{13}\frac{64\pi ^{2}}{N}\left( f_{0}+\frac{b_{0}}{2}%
(1-\frac{\gamma _{0}}{\beta _{0}})\right) =0\;,  \label{a14}
\end{equation}
namely
\begin{equation}
b_{0}=-\frac{4331}{396}\frac{N}{16\pi ^{2}}\;.  \label{a15}
\end{equation}
For the vacuum functional $\mathcal{E}\left( \widetilde{m}^{2}\right) $ one
gets
\begin{equation}
\mathcal{E=}\frac{9}{13}\frac{N^{2}-1}{N}\frac{1}{\left( g^{2}\right) ^{1-%
\frac{\gamma _{0}}{\beta _{0}}}}\left[ -\frac{\widetilde{m}^{4}}{2}+%
\widetilde{m}^{4}\frac{13}{3}\frac{Ng^{2}}{64\pi ^{2}}L\right] \;.
\label{a16}
\end{equation}
In terms of the scale independent variable $\widetilde{m}^{2}$, the gap
equation reads
\begin{equation}
\frac{\partial \mathcal{E}}{\partial \widetilde{m}^{2}}=0\;,  \label{a17}
\end{equation}
so that
\begin{equation}
-\widetilde{m}^{2}+\widetilde{m}^{2}\frac{26}{3}\frac{Ng^{2}}{64\pi ^{2}}L+%
\widetilde{m}^{2}\frac{13}{3}\frac{Ng^{2}}{64\pi ^{2}}=0\;.  \label{na1}
\end{equation}
Next to the solution, $\widetilde{m}^{2}=0$, we have the nontrivial solution
$\widetilde{m}_{sol}$ given by
\begin{equation}
-1+\frac{26}{3}\frac{Ng^{2}}{64\pi ^{2}}\log \left( \frac{\widetilde{m}%
_{sol}^{2}\left( g^{2}\right) ^{\frac{\gamma _{0}}{2\beta _{0}}}}{\overline{%
\mu }^{2}}\right) +\frac{13}{3}\frac{Ng^{2}}{64\pi ^{2}}=0\;.  \label{na2}
\end{equation}
In order to find a sensible solution of this equation, a suitable choice of
the scale $\overline{\mu }$ has to be done. Here, we take full advantage of
the RGE\ invariance of the vacuum functional $\mathcal{E}$, and set
\begin{equation}
\overline{\mu }^{2}=\widetilde{m}_{sol}^{2}\left( g^{2}\right) ^{\frac{%
\gamma _{0}}{2\beta _{0}}\text{ }}e^{-s}\;,  \label{na3}
\end{equation}
where $s$ is an arbitrary parameter which will be chosen at our best
convenience. The possibility of introducing this parameter relies on the
independence of the vacuum functional $\mathcal{E}$ from the renormalization
scale $\overline{\mu }$. Furthermore, recalling that
\begin{equation}
g^{2}(\overline{\mu })=\frac{1}{\beta _{0}\log \frac{\overline{\mu }^{2}}{%
\Lambda ^{2}}}\;,  \label{na4}
\end{equation}
and that, due to the change of the renormalization scheme,
\begin{equation}
\Lambda ^{2}=\Lambda _{\overline{MS}}^{2}e^{-\frac{b_{0}}{\beta _{0}}}\;,
\label{na5}
\end{equation}
for the effective coupling and the mass $\widetilde{m}_{sol}$, one finds
\begin{equation}
\left. \frac{Ng^{2}}{16\pi ^{2}}\right| _{1-loop}=\frac{12}{13}\frac{1}{%
(1+2s)}\;,  \label{a18}
\end{equation}
\begin{equation}
\left. \widetilde{m}_{sol}\right| _{1-loop}=\left( \frac{12}{13}\frac{16\pi
^{2}}{N}\frac{1}{(1+2s)}\right) ^{-\frac{\gamma _{0}}{4\beta _{0}}}e^{-\frac{%
b_{0}}{2\beta _{0}}}e^{\frac{13}{88}(1+2s)}e^{\frac{s}{2}}\Lambda _{%
\overline{MS}}\;.  \label{na18}
\end{equation}
Therefore, choosing $s=0.6$, and setting $N=3$, the following one-loop
estimates are found
\begin{equation}
\left. \frac{Ng^{2}}{16\pi ^{2}}\right| _{1-loop}\simeq 0.42\;,  \label{na19}
\end{equation}
\begin{eqnarray}
\left. \widetilde{m}_{sol}\right| _{1-loop} &\simeq &2.4\Lambda _{\overline{%
MS}}\simeq 560MeV\;,  \label{a19} \\
\Lambda _{\overline{MS}} &\simeq &233MeV\;,  \nonumber
\end{eqnarray}
\[
\left. \sqrt{\left\langle A_{\mu }^{a}A_{\mu }^{a}\right\rangle}
\right| _{1-loop}^{N=3}\simeq 0.22 GeV \;,
\]
and
\begin{equation}
\left. \mathcal{E}(\widetilde{m}_{sol})\right| _{1-loop}^{N=3}\simeq
-90\Lambda _{\overline{MS}}^{4}\simeq -0.265\left( GeV\right) ^{4}\;.
\label{na20}
\end{equation}
Note that the value obtained for $\widetilde{m}_{sol}$ is close to that
already reported for the dynamical gluon mass in the Landau gauge \cite
{Verschelde:2001ia,
Dudal:2003vv,Browne:2003uv,Browne:2004mk,Langfeld:2001cz, RuizArriola:2004en}%
. It should be remarked that the results $\left( \ref{na19}\right) $, $%
\left( \ref{a19}\right) $ have been obtained within a one-loop
approximation. As such, they can be taken only as a preliminary indication.
To find more reliable results, one has to go beyond the one-loop
approximation. Nevertheless, these calculations suggest that a non vanishing
gluon mass might emerge from the gap equation $\left( \ref{s6}\right) $.

\section{Conclusion}

\noindent In this work the issue of the dynamical mass generation
for gluons has been addressed. Due to color confinement, gluons
are not observed as free particles. Thanks to the asymptotic
freedom, the gauge field $A_{\mu }^{a}$ behaves almost freely at
very high energies, where we have a good understanding of its
properties. However, as the energy decreases the effects of
confinement cannot be neglected and it becomes more and more
difficult to have a clear understanding of $A_{\mu }^{a}$. As a
consequence, one does not exactly know what is the correct
starting point in the low energy region. We might thus adopt the
point of view of starting with a renormalizable action built up
with a gauge field $A_{\mu }^{a}$ which accommodates the largest
possible number of degrees of freedom. This would amount to take
as starting point a renormalizable massive action as considered,
for example, in expression $\left( \ref{s4}\right) $. The mass
parameter $m$ is not treated as a free parameter. Instead it is
determined by a gap equation, eq.$\left( \ref{s6}\right) $,
obtained by minimizing the vacuum functional
$\mathcal{E}$ of eq.$%
\left( \ref{s5}\right) $ with respect to the mass parameter $m$. A
preliminary analysis of this gap equation at one-loop shows that a
nonvanishing gluon mass might emerge. Also, the vacuum functional
$\mathcal{E}$ displays the important feature of obeying the
renormalization group equations.
\newline
\newline
Finally, we underline that the infrared behavior of the gluon
propagator is expected to be affected by several mass parameters,
with different origins. For instance, as pointed out in
\cite{Sobreiro:2004us,Dudal:2005na} in the case of the Landau
gauge, the gluon propagator turns out to be affected by both
dynamical gluon mass $m$ and Gribov parameter $\gamma$, which
arises from the restriction of the domain of integration in the
Feynman path integral up to the first Gribov horizon.  More
precisely, these parameters give rise to a three level gluon
propagator which exhibits infrared suppression
\cite{Sobreiro:2004us,Dudal:2005na}, namely
\begin{equation}
\left\langle A_{\mu}^a(k) A_{\nu}^b(-k) \right\rangle =
\delta^{ab} \left( \delta_{\mu\nu}-\frac{k_{\mu}k_{\nu}}{k^2}
\right) \frac{k^2}{k^4+m^2k^2+\gamma^4} \ .
\end{equation}

\section*{Acknowledgments.}

I am indebted to my friends and colleagues D. Dudal, J. A. Gracey,
and H. Verschelde for many valuable discussions. T. Turner and L.
R. de Freitas are gratefully acknowledged. We thank the Conselho
Nacional de Desenvolvimento Cient\'{\i}fico e Tecnol\'{o}gico
(CNPq-Brazil), the Faperj, Funda{\c c}{\~a}o de Amparo {\`a}
Pesquisa do Estado do Rio de Janeiro, the SR2-UERJ and the
Coordena{\c{c}}{\~{a}}o de Aperfei{\c{c}}oamento de Pessoal de
N{\'\i}vel Superior (CAPES) for financial support.

\end{document}